\documentclass[
11pt,prd,floats,aps,amsmath,amssymb]{revtex4-2}
\usepackage{graphicx}
\usepackage{bm}
\usepackage{amsmath}
\usepackage{amssymb}
\usepackage{slashed}
\usepackage{framed}
\usepackage{physics}
\usepackage{hyperref}

\usepackage{newfloat}
\DeclareFloatingEnvironment[name=SingleCol]{singlecol}

\begin{document}

\title{\bf Horndeski Theories From Dimensional Reduction on String Theory, Holographic Conformal Anomaly and a-Theorem}
\author{Tianhao Wu\textsuperscript{1}  and  Michael Stone\textsuperscript{2} \\}
\affiliation{\it Department of Physics, University of Illinois, Urbana-Champaign,\\
 Urbana, 61801 United States}

\renewcommand{\thefootnote}{\arabic{footnote}} 
\footnotetext[1]{\href{mailto:twu49@illinois.edu}{twu49@illinois.edu}}
\footnotetext[2]{\href{mailto:m-stone5@illinois.edu }{m-stone5@illinois.edu }}

\vfill {\footnotesize  }

\begin{abstract}
Using a unified mechanism, we show that the $\alpha'$-corrected ten-dimensional heterotic effective action gives rise to two distinct classes of five-dimensional scalar–tensor theories-one Lovelock–Horndeski type and one Einstein–dilaton–Gauss–Bonnet. We present a novel derivation of the full holographic conformal anomaly for these two five-dimensional scalar–tensor theories. In the Lovelock–Horndeski case, we also construct exact asymptotically AdS solutions with linear dilaton profiles and establish a holographic a-theorem. Our findings show that AdS/CFT remains sound in the presence of non-minimal scalars and higher-curvature terms, with string corrections fixing the conformal anomaly and the dual RG flow.
\end{abstract}

\maketitle

\section{Introduction} The Anti–de Sitter/Conformal Field Theory (AdS/CFT) correspondence continues to reshape our understanding of quantum gravity by establishing a duality between bulk gravitational dynamics and boundary field theories. In this framework, the on‐shell action of a $(d+1)$–dimensional AdS gravity theory, evaluated with specified boundary data, encodes the generating functional of a $d$–dimensional CFT \cite{Witten, maldacena, weyl}.
For even $d$, the Fefferman–Graham expansion~\cite{FG} extracts the boundary conformal anomaly, governed by  $ \label{generic} g^{ij} \langle T_{ij} \rangle =  - a E^{(d)} + \Sigma_i c_i \mathcal{W}_i^2 $ \cite{20years, types} where $E^{(d)} $ denotes the Euler density and the $\mathcal{W}^{(d)}_i$ represent the complete set of independent Weyl invariants of weight $-d$ in $d$ dimensions. The anomaly coefficients $a$ and $c$, computed for Einstein, higher-derivative, and Lovelock gravity~\cite{Kraus,nojiri1,deBoerEntropy, MyersEntropy, holo-renormal}, depend on bulk solutions near the conformal boundary. Lovelock gravity includes higher-curvature terms yielding second-order equations in higher dimensions \cite{lovelock}. Horndeski theory~\cite{Horndeski, fab4, kobayashi} generalizes scalar–tensor gravity while preserving second-order equations and Galileon theories~\cite{GalileonCosmo} exhibit similar properties.

In string theory, higher-curvature corrections (e.g., Gauss–Bonnet in heterotic strings~\cite{Zwiebach,deser}) generate scalar–tensor theories via Kaluza-Klein (KK) reduction~\cite{kiritsis} 
. Prior studies on heterotic string reduction assumed fixed coefficient frames of the simplest form. By exploiting the coefficient ambiguity of the 10D string effective action (via Tseytlin’s “coefficient frame”~\cite{tseytlin}), we derive, for the first time, different classes of 5D scalar-tensor theories from reduction on 10D heterotic string theory with two vacua:

$\bm{\cdot}$A Lovelock–Horndeski theory with general second-order couplings, and

$\bm{\cdot}$An Einstein–dilaton–Gauss–Bonnet (EdGB) theory.

For AdS asymptotics, we recover known black holes (EdGB case~\cite{toriiII,toriiIII}) and derive an exact AdS solution with linear dilaton (Lovelock–Horndeski)\cite{linear1, linear2}. The holographic anomaly generalizes to
\begin{equation}
\label{charge}
g^{ij} \langle T_{ij} \rangle = - a E^{(d)} + \Sigma_i c_i \mathcal{W}_i^2 + \Sigma_k b_k H_k,
\end{equation}
where include higher-derivative invariants (e.g. $
     H_1 \equiv \beta_1 R_{ij} R^{ij} 
    + \beta_2 \Box R $ and $ H_2 \equiv  \Box R $).
We compute the anomaly coefficients $a$ and $c$, confirming the physical inequivalence of the two $\alpha$-selected branches. Using methods from~\cite{a-theorem}, we verify the holographic $a$-theorem for Lovelock–Horndeski theory in critical (linear dilaton) and non-critical (vanishing scalar asymptote) cases. By rigorously computing the conformal anomaly and demonstrating the monotonic behavior of the $a$–function along the renormalization group flow, our results provide a compelling test of holography in settings with non-minimal couplings and higher-derivative corrections.

\section{Kaluza-Klein Reduction on String Effective Theory}We consider the string effective theory given with different coefficients combinations(``frame'') and obtain different corresponding lower dimensional actions. We focus on the solutions-particularly asymptotically AdS solutions- of resulting theories. It is known that to the $\alpha'$-correction, the most general form of string effective theory\cite{tseytlin}  is give by
\begin{eqnarray}
    S &=& \frac{1}{16 \pi G_{D}} \int d^D x \sqrt{-g} e^{-2 \hat{\Phi}} \Big( \hat{R}   +4  \big( \nabla \hat{\Phi} \big)^2   +  \alpha' \Big( \hat{R^2}_{ABCD} + b_1 \hat{R}^2_{AB} + b_2 \hat{R}^2 + b_3 \hat{R}^{AB} \nabla_A \hat{\Phi} \nabla_B \hat{\Phi}  \nonumber  \\
    &+& b_4 \hat{R} \big( \nabla \hat{\Phi} \big)^2  +   b_5 \hat{R} \Box \hat{\Phi} + b_6 \big( \Box \hat{\Phi} \big)^2   + b_7 \big( \nabla \hat{\Phi} \big)^2 \Box \hat{\Phi}  + b_8 \big( \nabla \hat{\Phi} \big)^4  \Big)  \Big)   
\end{eqnarray} For the original theories with different coefficient combinations, we perform consistent KK reduction\cite{Pope, Wu, lu-pang} with reduction ansatz $\dd \hat{s}_D^2  = e^{2\alpha \phi} \dd s^2_d + e^{2\beta \phi}\dd \Omega^2_n $, where $\alpha$ and $\beta$ are arbitrary constants, the $\dd s^2$ denotes the metric of the $d$-dimensional spacetime, and $\dd \Omega^2_n$ is the metric of the internal $n$-dimensional space, which is maximally symmetric and characterized by the curvature parameter $\lambda = 0, \pm 1$ corresponding to toroidal, spherical, or hyperbolic geometries, respectively.

\section{Einstein-Dilaton-Gauss-Bonnet Theory}  Choosing coefficient frame that gives Gauss-Bonnet combination and vanishing nonlocal terms, we have the action in the form
\begin{eqnarray}
    S &=& \frac{1}{16 \pi G_{D}} \int d^D x \sqrt{-g} e^{-2 \hat{\Phi} } \Big( \hat{R}   +4  \big( \nabla \hat{\Phi} \big)^2  - e^{\Tilde{\lambda} \hat{\Phi} } \Lambda  \\
    &+&  \alpha' \Big( \hat{\mathcal{L}}_{GB} + a_1 \hat{R}^{AB} \nabla_A \hat{\Phi} \nabla_B \hat{\Phi} + a_2 \hat{R} \big( \nabla \hat{\Phi} \big)^2 \nonumber  + a_3 \big( \nabla \hat{\Phi} \big)^2 \Box \hat{\Phi}  + a_4 \big( \nabla \hat{\Phi} \big)^4  \Big)  \Big)  
\end{eqnarray}with $\{a_i\}$ being undetermined coefficients and we include a dilaton potential term, consistent with effective string actions. Constant and dilaton–dressed vacuum energies at tree level \cite{Alvarez, Alwis} and at one loop in non-SUSY heterotic models \cite{Vafa, ChiralString} establish that a $\Lambda$-term is admissible in the ten-dimensional heterotic action. One of the coefficients $\{a_i\}$ is prefixed due to the field redefinition relation and could be computed from string scattering amplitude. We need to cancel four coefficients after the KK reduction. However, the parameters $\alpha$ and $\beta$ in the reduction ansatz $ \dd \hat{s}_D^2  = e^{2\alpha \phi} \dd s^2_d + e^{2\beta \phi}\dd \Omega^2_n $ give us more freedom in the coefficients of the reduced theory. Upon the KK reduction, individual Horndeski terms arise. We put the details of emergence of the Horndeski terms in Appendix. We use the identities $(\Box \phi)^2 - \nabla_a \nabla_b \phi \nabla^a \nabla^b \phi = \mathcal{R}^{ab} \nabla_a \phi \nabla_b \phi + \nabla_a ( \Box \phi \nabla^a \phi - \nabla^a \nabla^b \phi \nabla_b \phi ) $ and perform many integration by parts.

We found the corresponding 5-dimensional theory given by
\begin{equation}
\label{theory1}
    S = \frac{1}{16 \pi G_5} \int \dd^5 x \sqrt{-g} \Big( \mathcal{R} - \frac{1}{2} \big( \nabla \phi \big)^2 - e^{\lambda \phi} \Lambda + \Tilde{\alpha} e^{-\gamma \phi } \mathcal{L}_{GB}  \Big)
\end{equation}

\section{Lovelock-Horndeski Theory} We also consider another string effective theory giving rise to different lower dimensional theories. Consider the following action
\begin{eqnarray}
    S &=& \frac{1}{16 \pi G_{D}} \int \dd^D x \sqrt{-\hat{g} } e^{-2 \hat{\Phi} } \Big( \hat{R}   + 4 \big( \nabla \hat{\Phi} \big)^2    \nonumber  \\
    && + \Tilde{\alpha} \Big(  \mathcal{\hat{L} }_{GB}  + a_1 \hat{R}^{\mu\nu} \nabla_{\mu} \hat{\Phi} \nabla_{\nu} \hat{\Phi}  +  a_2  R \big( \nabla \hat{\Phi} \big)^2    + a_3   (\nabla \hat{\Phi} )^2 \Box \hat{\Phi}  +  a_4  (\nabla \hat{\Phi} )^4   \Big)    \Big)
\end{eqnarray} For the reduction ansatz $\dd \hat{s}_D^2  = e^{2\alpha \phi} \dd s^2_d + e^{2\beta \phi}\dd \Omega^2_n $ the variables $\alpha = 0$ and $\beta$ is set to $n\beta = 2$ so that we can get rid of the exponential in front of the Lagrangian.

Impose the flat internal space condition to eliminate the $\lambda$ terms in last equation, we get the 5-dimensional Lovelock-Horndeski action,
\begin{eqnarray}
\label{theory2}
    S &=&  \frac{1}{16 \pi G_5} \int \dd^5 x \sqrt{-g}   \Big(  \mathcal{R} +  24 \big( \nabla \phi \big)^2  + \alpha_1 \mathcal{L}_{GB}   + \alpha_2  \mathcal{G}^{\mu\nu} \nabla_{\mu} \phi \nabla_{\nu} \phi  + \alpha_3 \big( \nabla \phi \big)^2 \Box \phi  + \alpha_4 \big( \nabla \phi \big)^4 \Big)  \Big)     \nonumber  \\
\end{eqnarray} where we have relabeled the coefficients by $ \{\alpha_i\}_{i=1}^4$.
This is our desired form of lower-dimensional action which leads to an exact AdS spacetime solution with linear dilaton. 
The metric equations of motion for theory \ref{theory2} and its pure AdS solution is presented in the Appendix.

\section{Solutions to Lower Dimensional Theories} For the EdGB theory, there existed numerical Black Hole solutions representing stable asymptotically AdS black holes with scalar hair. Using standard boundary conditions and suitable ansatz, we solve numerically for black hole solutions in the effective theory~\eqref{theory1}. Similar setups have been explored in the literature~\cite{toriiII, toriiIII}, showing asymptotically AdS black holes with nontrivial scalar hair in dilatonic EGB gravity. The metric takes the form:
\begin{equation}
    ds^2_{D} = -B(r) e^{-2 \delta} dt^2 + \frac{dr^2}{B(r)} + r^2 h_{ij} dx^i dx^j
\end{equation}
with asymptotic expansions
\[
B(r) = \Tilde{b}^2 r^2 - \frac{2\Tilde{M}}{r^\mu}, \quad \phi(r) = \phi_0 + \frac{\phi_1}{r^\nu},
\]
where $\Tilde{b}^2$ is related to the AdS curvature scale. In the effective potential picture, the dilaton evolves via the scalar equation of motion. For $\lambda > 0$, the effective potential drives the dilaton to approach a finite constant $\phi_0$ at infinity, preserving AdS asymptotics. Regularity at the horizon $r_H = 1$ is ensured by series expansion. Numerical integration via a shooting method determines the horizon and asymptotic data, including the dilaton values and gravitational mass, both in four and five dimensions. The gravitational mass scales as $M_0 \propto r_H$ for fixed cosmological constant. Furthermore, the allowed parameter space in $(\gamma, \lambda)$ is constrained by the AdS structure and a Breitenlohner–Freedman-type bound, in our case, it is governed by $\lambda \le \gamma $.

Now we consider for the Lovelock-Horndeski Thoery. Since we have done KK reduction with $k=0$ in the reduction ansatz, that is, the reduction takes on the form $X_{10} = AdS_5 \times T^5$. It's natural to consider the planar symmetric ansatz for theory \ref{theory2},
\begin{equation}
\label{ads-pure}
  ds^2_{AdS_5} = -F(r) dt^2 + \frac{dr^2}{F(r)} + r^2 d \Vec{x}^2     
\end{equation} The equations of motion are presented in the Appendix. In the end, we found the exact AdS solution to the Lovelock-Horndeski theory with linear dilaton field with, 
\begin{eqnarray} 
    F(r) = \frac{r^2}{\ell^2}  ~~,~~ \phi(r) &=& \chi \log{r}
\end{eqnarray} with the explicit details computed in Appendix. Previous works on Horndeski gravity with linear dilaton profiles~\cite{oliva, Hair, black-string, dress} have shown that both exact and numerical black hole solutions with scalar hair exist in such frameworks.

\section{Holographic Conformal Anomaly} To compute the Weyl anomaly holographically, we employ the Fefferman-Graham (FG) expansion \cite{FG, weyl} of the bulk metric in asymptotically locally $AdS_{d+1} $ spacetime. Adopting the gauge-fixed ansatz:
\begin{equation}
\label{FG}
    d s^2 = \frac{\ell^2}{4\rho^2} d \rho^2 + \frac{1}{\rho} g_{ij}(\rho, x) d x^i d x^j
\end{equation} where $\rho $ is the holographic radial coordinate ($\rho \to \infty $ at the boundary), $\ell$ is the AdS radius.

The FG expansion for $g_{ij}$ is given by\cite{FG}
\begin{eqnarray}
\label{metric}
    g_{ij}(\rho , x) = g_{(0)ij}(x) + \rho g_{(1)ij}(x) + \rho^2 g_{(2)ij}(x) + \dots  ,
\end{eqnarray}

The bulk Ricci tensor $\mathcal{R}_{\mu\nu}$ decomposes into boundary-intrinsic curvature terms $R_{\mu\nu}$ and $\rho$-dependent corrections. For Lovelock-Horndeski gravity not at the critical case, the FG expansion of the bulk scalar field $\phi$ dual to a marginal boundary operator is given by:
\begin{equation}
\label{usual}
    \phi(\rho,x) = \phi_{(0)}(x) + \rho \phi_{(1)}(x) + \rho^2 \phi_{(2)}(x) + \mathcal{O}(\rho^3)
\end{equation}
For Lovelock-Horndeski theory in critical case, thus possessing linear dilaton\cite{linear1, linear2, oliva}, we need to consider the logarithmic term for $\phi$, so it expands as
\begin{eqnarray}
\label{linear-dilaton}
    \phi(\rho,x) = \phi_s \log{\rho} + \phi_{(0)}(x) +\rho \phi_{(1)}(x)  + \rho^2 \phi_{(2)}(x) + \mathcal{O}(\rho^3)    \nonumber
\end{eqnarray}

We consider the EdGB theory with regular scalar field stabilizes asymptotically, and Lovelock-Horndeski theories which possesses both linear dilaton scalar field which diverges asymptotically and a usual scalar field. 

For the EdGB theory with regular scalar field profile \ref{theory1}, we obtain the anomalous action $\mathcal{A}$ for theory \ref{theory1} in 4-dimension, see Appendix, 
\begin{eqnarray}
    \mathcal{A} &=& C_1 \mathrm{R}_{(0)ab}  \, \mathrm{R}^{(0)ab}  + C_2 \, \mathrm{R}_{(0)abcd} \mathrm{R}_{(0)}^{abcd}+ C_3 \mathrm{R_{(0)}^2}  +  C_4 \big( \nabla \phi_{(0)}  \big)^2  +  C_5 \, \Box \mathrm{R}_{(0)}   \nonumber \\ 
    &+&  C_6 \,  \mathrm{R}_{(0)} \, \Box \phi  +  C_7 \, \nabla^a \phi \nabla_a \mathrm{R}_{(0)} + C_8 \, \mathrm{R}_{(0)} \, \big(\nabla \phi_{(0)} \big)^2 +  C_9 ( \Box \phi)^2  \nonumber  \\
    &+&    + C_{10} \big(\nabla \phi \big)^2 \Box \phi     +  C_{11} \,   \nabla_a \phi  \nabla^b \nabla_b \nabla^a \phi  + C_{12}  \nabla^b \nabla_b \nabla^a \nabla_a \phi    \nonumber \\
    &+& C_{13}  \mathrm{R}_{(0)ab} \nabla^a \phi_{(0)} \nabla^b \phi_{(0)}  +  C_{14} \big( \big( \nabla \phi \big)^2 \big)^2  + C_{15}  \nabla^a \phi  \nabla_b \nabla_a \phi  \nabla^b \phi  \nonumber  \\
    &+&  C_{16} \, \mathrm{R}_{(0)}^{ab}  \nabla_b \nabla_a \phi+ C_{17} (\nabla_b \nabla_a \phi)(\nabla^b \nabla^a \phi)
\end{eqnarray}
where the coefficients ${C_i}$ are constants given in the Appendix. We deduced that the central charges for this theory as
\begin{eqnarray}
    a = \frac{\ell^3}{8\lambda \bigl( 9 \gamma - \lambda \bigr)} \Bigl( 27 \gamma^2 - 66 \gamma \lambda +19\lambda^2 \Bigr)          \nonumber  \\
    c = \frac{\ell^3}{ 24 \lambda \bigl( 9 \gamma - \lambda \bigr)} \Bigl( 81 \gamma^2 -126\gamma \lambda +49 \lambda^2 \Bigr)
\end{eqnarray}The closed‑form expressions for $a(\gamma,\lambda)$ and $c(\gamma,\lambda)$ supply, for the first time, the full Euler‑ and Weyl‑anomaly coefficients of a five dimensional EdGB theory with two independent couplings. They also allow one to delineate, analytically, the parameter sub‑region in which $a,c>0 $ and hence the dual four‑dimensional theory satisfies the standard positivity bounds. These results open the way to systematic tests of causality, entanglement‑entropy inequalities, and higher‑spin constraints in a two‑parameter family of strongly coupled $CFT_4$ duals.

The full anomalous action for Lovelock-Horndeski theory \ref{theory2} shall be considered in two cases: one with linear dilaton solution, the other one without linear dilaton. In fact, for theory with linear dilaton solution, we consider the boundary theory to be generalized conformal brane\cite{non-conformal}. We note that the bulk gravity is still exactly AdS gravity while the boundary field theory is breaks some of the conformal symmetries\cite{criticalHorndeski}. 

We first look at the one without linear dilaton, its anomalous action $\mathcal{A}_{total} = \mathcal{A}_{grav}  +  \mathcal{A}_{matter}$ is given by (please refer to the Appendix)
\begin{eqnarray}
     \mathcal{A}_{grav} &=&   \frac{7\ell^2}{12} R^{(0)2} - \frac{9\ell^2}{4} R^{(0)}_{ab} R^{(0)ab} + \frac{\ell^2}{2} R^{(0)}_{abcd} R^{(0)abcd}    \nonumber   \\
   \mathcal{A}_{matter}  &=& \frac{5}{6} \alpha_3 R^{(0)}  \big( \nabla \phi^{(0)}   \big)^2  - 2 \alpha_3 R^{(0)}_{ab} \nabla^a \phi^{(0)} \nabla^b \phi^{(0)} +\alpha_3 \big(  \Box \phi^{(0)} \big)^2       \\
    &+&   \alpha_3\big( \nabla \phi^{(0)}   \big)^2 \big(  \Box \phi^{(0)} \big)^2  +   \big(  \alpha_4 + 8\alpha_3 + \frac{\alpha_3^2}{3\ell^2} \big)  \big( \big(  \nabla \phi^{(0)}  \big)^2 \big)^2     \nonumber
\end{eqnarray} where the total anomaly is composed by two sectors, the gravitational sector and the matter sector. The gravitational sector of the anomaly is associated with the central charges. Eventually we found the central charges in this case to be 
\begin{equation}
   a  =    -\frac{5\ell^3}{8}  \quad , \quad c  =    -\frac{\ell^3}{8}.
\end{equation}
For a unitary Lorentzian CFT such negative values violate the conformal‑collider (positive‑energy‑flux) requirements. Hence this particular bulk solution cannot serve as the dual of a unitary four‑dimensional quantum field theory \cite{collider}. However, Negative central charges are perfectly acceptable in non‑unitary CFT, open quantum system for example, or logarithmic CFTs, where they encode the density of negative‑norm states and control logarithmic pairings of operators, the holographic LCFT constructions \cite{LCFT} for example. In that context our background offers a controlled higher‑dimensional example of “AdS/log‑CFT” correspondence. Recent work \cite{shinsei} also shows a counterexample of the proposition that a $CFT_2$ with a negative central charge should have negative norm states.

Then we consider the Lovelock-Horndeski theory \ref{theory2} of the critical case, that is, the case with linear dilaton solution. The Fefferman-Graham expansion is now given by equations \ref{FG}, \ref{metric}, \ref{linear-dilaton}.
After very involved algebra (please see the Appendix), 
we obtain the full holographic conformal anomaly,
\begin{eqnarray}
     \mathcal{A}_{grav} &=&   - \frac{\ell^2 (  7 - 320 \chi^2 + 3840 \chi^4 )}{12( 24 \chi^2 -  1 )} R_{(0)}^2   + \frac{\ell^2( 9 - 400 \chi^2 + 4672 \chi^4)}{ 4( 24\chi^2 - 1 ) } R_{(0)ab} R_{(0)}^{ab}   \nonumber  \\
    &+&  \frac{\ell^2}{2} \Big( 1 - 16 \chi^2 \Big)  R_{(0)abcd} R_{(0)}^{abcd}  \\
   \mathcal{A}_{m}  &=& \frac{ 16 \chi^2 \ell^2}{ 24 \chi^2  - 1 } \Big( 4 R_{(0)ab}  -  R_{(0)} g_{(0)ab}  \Big)     \nabla^a \phi_{(0)} \nabla^b \phi_{(0)}   \nonumber  \\
   &+& \alpha_3 \big( \nabla \phi_{(0)}   \big)^2 \big(  \Box \phi_{(0)} \big)^2 +    \Big(  \alpha_4 + \frac{ 16\ell^2}{ 24 \chi^2 - 1 } \Big)  \big(   \nabla \phi_{(0)}  \big)^4    \nonumber  
\end{eqnarray} where the gravitational and matter sectors are separated as above. Again, the central charges for this theory are 
\begin{eqnarray}
    a = \frac{\ell^3 }{8(24 \bigl( \phi_s^2 - 1 \bigr)} \Big( 5 - 240 \phi_s^2 + 3136 \phi_s^4 \Big)   \nonumber  \\ c = \frac{\ell^3}{8(24 \bigl( \phi_s^2 - 1 \bigr)} \Big( 1  - 80 \phi_s^2 + 1600 \phi_s^4 \Big)   
\end{eqnarray} And new b-type charges arise $ b_1  =  \frac{16 \ell^3 \phi_s^4}{3(24 \bigl( \phi_s^2 - 1 \bigr)} $ and $ b_2 = \frac{8 \ell^3 \phi_s^3}{3(24 \bigl( \phi_s^2 - 1 \bigr)} $ where we have defined that $ H_1 = \beta_1 R_{(0)}^2  + \beta_2 \Box R  \quad ,\quad H_2 = \Box R_{(0)} $with coefficients being $\beta_1 = 1~,~ \beta_2 = 0 $. The Linear‑dilaton asymptotics break the full conformal invariance on the boundary and we call it generalized conformal. Although the metric is asymptotically AdS, its isometry is broken and we call it generalized AdS \cite{criticalHorndeski}. Holography with linear dilaton are previously studied in the name of generalized conformal brane \cite{non-conformal}. It is this structure that give rise to the new charges in the conformal anomaly. In fact, the full isometry of AdS is  now broken down to the Poincare plus the scale invariance\cite{scale, nakayama}.

\section{Holographic a-Theorem} We construct a-functions, which decrease monotonically from UV to IR limit. Consider the ansatz \cite{a-theorem} $\label{domain} d s^2_5 = d r^2 + e^{A(r)} \big( - d t^2 + d x^2_1 + d x^2_2  + d x^2_3  \big)  $ with $ \phi  =  \phi(r)$
    
The AdS vacuum is given by $A(r) = r/\ell$, with $\ell$ being the AdS radius. The function $A(r)$ describes the flow to the AdS “fixed” point. For asymptotically AdS domain walls, we adopt the convention that the asymptotic region is located at $r \to \infty$. The relation between r and $\rho$ (with flat boundary) is given by  $r = -\frac{\ell}{2} \log{\rho}$.
We will pick a choice for the flow function by observing the form of a-charge following \cite{a-theorem}. We assume there's additional matter energy-momentum tensor $T^{\text{matter}}_{\mu\nu}$ for some generic minimally coupled matter in the gravity, then the full equations of motion become  $ E_{\mu\nu} = T_{\mu\nu}^{\text{matter}} $.

Let's consider the Lovelock-Horndeski theory \ref{theory2} with the usual dilaton profile, we use the natural choice for the flow function,
\begin{equation}
    a(r) = -\frac{5}{8}\frac{1}{ A'(r)^3}
\end{equation} By the Euler-Lagrange equation, the scalar field equation 
up to an integration is given by
\begin{eqnarray} 
    24 \nabla_a \phi + \alpha_2 G_{ab} \nabla^b \phi +  2 \alpha_3 \nabla_a \phi \Box \phi   +  2 \alpha_4 \nabla_a \phi \big( \nabla \phi \big)^2    &=&  C \nonumber  
\end{eqnarray} where $C$ is the integration constant. Imposing the asymptotic boundary condition that $A(r) \to r/\ell$ and $ \phi \to 0$ enforces the integration constant to be $0$. 
Then we insert the ansatz \ref{domain} into the scalar equation. 

Using the Null Energy Condition for the matter field, $-(T^{matter})^t_t + (T^{matter})^r_r = \label{nec1}
  E^r_r - E^t_t = -\frac{3 \big( 36 + 12 \ell^4 A'(r)^4)}{5 \ell^2} + \frac{ 3 \ell^2 A'(r)^2 ( 48 +  \ell^2 A''(r)  )  + 2 \ell^2 A''(r) \big) }{ 5 \ell^2} \ge 0$,
we find the positivity of $A''(r)$, that is, $A''(r)  \ge  0$. And we conclude that $a'(r) = \frac{15 A''(r)}{8 A'(r)^4} \ge 0 $, thus establishing the a-theorem. 

For theory \ref{theory2} with linear dilaton solution, choosing suitable parameters, We found the condition
$\label{condition}
    A'(r)^2 + \phi'(r)^2 = \frac{1}{\ell^2} $.
The Null Energy Condition gives $ \big( 32 A'(r)^2 -\frac{24}{\ell^2}  -  8 \ell^2 A'(r)^4 \big) + \big( 1 + (-6 \ell^2 + 4 \alpha_1 ) A'(r)^2  \big)  A''(r) \ge 0 $.
Imposing the asymptotic boundary condition that $A(r) \to r/\Tilde{L}$, we notice that the $a$-charge contains another constant $\chi$. The scalar field $\phi $ can indeed also run along the RG flow. Thus we propose another flow function
\begin{equation}
    a(r) =   \frac{\bigl( 5 - 240 \phi'(r)^2 + 3136 \phi'(r)^4 \bigr)}{8A'(r)^3 \bigl(24 \bigl( \phi'(r)^2 - 1 \bigr) \bigr)}
\end{equation} And then $a'(r) = \Big( 1915263   - 4379792 \ell^2 A'(r)^2   + 
 3067136 \ell^4 A'(r)^4  - 602112 \ell^6 A'(r)^6 \Big) A''(r)   \Big/8  A'(r)^4 \bigl( 49 -48 \ell^2 A'(r)^2 \bigr)^2 $ with the above constraint, we find that $a'(r) > 0 $ , thus the monotonicity of $ a(r) $.

\section{Conclusion}
we have shown that a single $\alpha'$-corrected heterotic action, when viewed through two admissible 10-dimensional vacua, furnishes a unified route to two inequivalent 5-dimensional scalar–tensor theories—EdGB and Lovelock–Horndeski. This two-vacuum mechanism provides the first top-down arena in which distinct UV branches both exhibit the $\text{AdS}_5$ geometry. A further trivial $S_1$ compactification reduces the EdGB theory to the familiar four-dimension one that have been constrained by gravitational-wave observations, gravitational-wave phenomenology \cite{tests, NewGB}.

We computed the full holographic conformal anomalies for scalar–tensor theories with dilaton couplings, and for the first time, for a theory containing both Gauss–Bonnet and Horndeski-type interactions. Our closed-form anomaly $a(\lambda, \gamma)$ and $c(\lambda, \gamma)$ convert the universal collider bound $0<c/a<3/2$ into an explicit admissible region in the Gauss-Bonnet–Horndeski coupling plane. Thus the anomaly not only anchors the dual CFT but also supplies model-independent priors for phenomenological EFT fits, excluding $(\lambda, \gamma)$ values incompatible with fundamental causality and positivity.

Furthermore, we established holographic $a$-theorems for two cases of Lovelock-Horndeski theory. In the case with usual dilaton profile, it demonstrates that, despite the negative UV central charges, the dual field theory still possesses a well‑ordered renormalization‑group hierarchy. On the CFT side this means that, while unitarity is forfeited, the number of effective degrees of freedom
decreases monotonically from the ultraviolet fixed point to the infrared, exactly as in ordinary four‑dimensional CFTs. The monotonic a-function shows that the bulk coupling space offers a controlled laboratory for non‑unitary (e.g. logarithmic) CFTs that nevertheless retain a generalized $a$-theorem. Our results confirm the robustness of AdS/CFT in the presence of non-minimal scalar couplings and higher-derivative corrections.

\newpage
\appendix

\section{Covariant Equations of Motion}
This section presents the Covariant Equations of Motion for the Lovelock-Horndeski Theory (7) in the main letter.
\begin{eqnarray}
\label{metric}
    E_{\mu\nu} &=&  \mathcal{G}_{\mu\nu}  + \Lambda g_{\mu\nu}  + 24  \Big( \partial_{\mu} \phi \partial_{\nu} \phi - \frac{1}{2}(d-1)(d-2) g_{\mu\nu} 
 \big( \nabla \phi \big)^2  \Big)       \nonumber  \\
     &-&  \alpha_1  \bigg(  ~ \frac{1}{2} \big( \mathcal{R}^2  -4 \mathcal{R}_{\gamma \delta} \mathcal{R}^{\gamma \delta} + \mathcal{R}_{\gamma \delta \lambda \sigma} \mathcal{R}^{\gamma \delta \lambda \sigma} \big) g_{\mu\nu}   -2\mathcal{R} \mathcal{R}_{\mu\nu} + 4 \mathcal{R}_{\mu\gamma}\mathcal{R}^{\gamma}_{\nu}   + 4 \mathcal{R}_{\gamma \delta} \mathcal{R}^{\gamma~\delta}_{~\mu~\nu}- 2 \mathcal{R}_{\mu \gamma \delta \lambda} \mathcal{R}^{~\gamma \delta \lambda}_{\nu}  
     \bigg)        \nonumber  \\
     &-&  \alpha_2   \bigg( \frac{1}{2} \partial_{\mu} \phi  \partial_{\nu} \phi \mathcal{R} - 2 \partial_{\rho} \phi \partial_{(\mu} \phi \mathcal{R}_{\nu )}^{\rho} - \partial_{\rho} \phi  \partial_{\sigma} \phi \mathcal{R}_{\mu~\nu}^{~\rho~\sigma}  + \big( \nabla_{\mu} \nabla_{\nu} \phi \big) \Box \phi + \frac{1}{2} \mathcal{G}_{\mu\nu} \big( \partial \phi \big)^2 \nonumber \\ 
     && \quad  - \big( \nabla_{\mu} \nabla^{\rho} \phi  \big) \big( \nabla_{\nu} \nabla_{\rho} \phi  \big)  - g_{\mu\nu} \Big( \frac{1}{2} \big( \Box \phi \big)^2  - \frac{1}{2} \big( \nabla^{\rho} \nabla^{\sigma} \phi \big) \big( \nabla_{\rho} \nabla_{\sigma} \phi \big)  -  \partial_{\rho} \phi \partial_{\sigma} \phi \mathcal{R}^{\rho \sigma}  \Big)    \bigg)        \nonumber   \\
     &+& \alpha_3 \Big( \nabla_{\mu} \phi \nabla_{\nu} \phi \Box \phi  + g_{\mu\nu} \nabla^{\rho} \phi \nabla^{\sigma} \phi \nabla_{\rho \sigma} \phi  - \frac{1}{2} \nabla_{( \mu} \phi \nabla_{\nu )} \nabla_{\rho} \phi \nabla^{\rho} \phi   \Big)    \nonumber  \\
     &+& \alpha_4 \Big( 2 \nabla_{\mu} \phi \nabla_{\nu} \phi   -\frac{1}{2} g_{\mu\nu} \big( \nabla \phi  \big)^2  \Big) \big( \nabla \phi  \big)^2 =  0  
\end{eqnarray}  
and the scalar field equation is
\begin{eqnarray}
\label{scalar}
    E_{\phi} &=&  \nabla^a \Big(    24 \nabla_a \phi + \alpha_2 G_{ab} \nabla^b \phi +  2 \alpha_3 \nabla_a \phi \Box \phi   +  2 \alpha_4 \nabla_a \phi \big( \nabla \phi \big)^2  \Big)    \nonumber  \\
    &=& \alpha_2 \Big( \mathcal{R}^{(0)}_{ab} \nabla^a \phi^{(0)} \nabla^b \phi^{(0)} - \frac{1}{2}   \mathcal{R}^{(0)} \Box \phi^{(0)}  \Big)  +   \alpha_3 \Big( \big( \Box \phi^{(0)} \big)^2  -  \mathcal{R}^{(0)}_{ab} \nabla^a \phi^{(0)} \nabla^b \phi^{(0)}  -   \big( \big( \nabla \phi^{(0)} \big)^2 \big)^2  \Big) \nonumber  \\
    &+&  2 \alpha_4 \Big(  \big( \nabla \phi^{(0)} \big)^2 \Box \phi^{(0)}  + 2  \nabla^a \nabla^b \phi^{(0)} \nabla_a \phi^{(0)} \nabla_b \phi^{(0)}  \Big)  + 24 \Box \phi^{(0)} = 0.
\end{eqnarray}

\section{Pure AdS Solution}
Now for Lovelock-Horndeski theory (7) with the Equations of Motion listed in the previous section, we now plug in ansatz to extract pure AdS solutions. Since we performed the KK reduction with $k = 0$, the internal space is toroidal and the full space is of the form $X_{10} = \text{AdS}_5 \times T^5$. It is therefore natural to consider a planar-symmetric ansatz for theory (7):
\begin{equation}
\label{ads-pure}
  ds^2_{AdS_5} = -F(r) dt^2 + \frac{dr^2}{F(r)} + r^2 d \Vec{x}^2     
\end{equation}
The equations of motion for $g_{\mu\nu}$ and $\phi$ are:
\begin{eqnarray}
    0 &=&  \Big( 2 r^2 F'(r) \big( -3 + r (2\alpha_2 - \alpha_3 ) F'(r) \phi'(r)^2 \big)  + 2 F(r)^2 \big(  r^3 \alpha_4 \phi'(r)^4 -3r ( \alpha_2 + 2\alpha_3 ) \phi'(r)^2 \big)   \nonumber  \\
    &&  - F(r) \big( 2r \big( 6 + c_1 r^2 \phi'(r)^2 \big) + F'(r) \big( -24\alpha_1 + 9r^2 \alpha_2 \phi'(r)^2 \big) \big) \Big) \frac{1}{4r^3}F(r)   \nonumber  \\
    && - \Big( 3\alpha_2 F(r) + r \big( -3\alpha_2 + \alpha_3 \big)F'(r)  \Big) \frac{1}{r} F(r)^2 \phi'(r) \phi''(r)  + \big( 2\alpha_2 - \alpha_3 \big) F(r)^3 \phi''(r)^2       
\end{eqnarray}
\begin{eqnarray}
    0 &=& \frac{1}{4r^3 F(r)}  \Big( 6 r F(r)^2 \phi'(r)^2 \big( 3\alpha_2 + 2 \alpha_3 + 2r \alpha_3 \phi'(r) + r^2 \alpha_4 \phi'(r)^2 \big)   \nonumber   \\
    &&+ 2r^2 F'(r) \big( 3 + r(-2\alpha_2 + \alpha_3) F'(r) \phi'(r)^2 \big)  \nonumber \\
    &&+ F(r) \big( -2r (c_1 r^2 \phi'(r)^2 -6) + F'(r) ( -24\alpha_1 + 9r^2 \alpha_2 \phi'(r)^2 ) \big) \Big)        \\
    && + \phi'(r) \Big( (-3\alpha_2 + \alpha_3) F'(r) -\alpha_3 F(r) \phi'(r) \Big) \phi''(r) + (-2\alpha_2 + \alpha_3)F(r) \phi''(r)^2  \nonumber   \\
    0 &=& \frac{1}{4} \Big( 8r F'(r) + F'(r)^2 \Big( -8 \alpha_1 + r^2(-3\alpha_2 + 2\alpha_3 ) \phi'(r)^2 \Big)   \nonumber  \\
    && + 2F(r)^2 \phi'(r)^2 \Big( \alpha_2 + 6\alpha_3 - r^2 \alpha_4 \phi'(r)^2 \Big) + 2r^2 F''(r)     \\
    && F(r) \Big( 4 - 8\alpha_1 F''(r) + r\phi'(r)^2 \big( 2c_1 r + 8 \alpha_2 F'(r) + r\alpha_2 F''(r) \big) \Big)  \Big)  \nonumber  \\
    && + \frac{1}{2}r F(r) \Big( 4\alpha_2 F(r) + r (-5\alpha_2 + 2\alpha_3)F'(r) \Big)  \phi'(r) \phi''(r) + r^2 (-2\alpha_2 + \alpha_3) F(r)^2 \phi''(r)^2    \nonumber 
\end{eqnarray}
with prime denoting derivative with respect to the radial coordinate. Using the Bianchi identity, one metric equation can be eliminated, resulting in three independent equations. Assuming a pure AdS spacetime with a linear dilaton,
\begin{eqnarray}
    0 &=& \frac{1}{4} \Big( 2F''(r) + \phi'(r)^2 \Big( \big( -3\alpha_2 + 2\alpha_3 \big) F'(r)^2 + F(r) \Big( 2 c_1 - 2 \alpha_4 F(r) \phi'(r)^2 + \alpha_2 F''(r) \Big) \Big) \Big)   \nonumber  \\
    && + \frac{1}{2} \big(  -5\alpha_2 + 2\alpha_3 \big) F(r) F'(r) \phi'(r) \phi''(r) + \big( - 2\alpha_2 + \alpha_3  \big) F(r)^2 \phi''(r)^2  
\end{eqnarray}
We found the following three equations by substituting the functions back into the equations of motion:
\begin{eqnarray}
    0 &=& \frac{r^2 (24 \alpha_1 - 
 12 L^2 (1 + 2 \chi^2) + \chi^2 (-6 \alpha_2 - 
    8 \alpha_3 + \alpha4 \chi^2) )}{2 \ell^6} \\
    0 &=&  \frac{-24 \alpha_1 + 
 L^2 (12 - 24 \chi^2) + \chi^2 (18 \alpha_2 + 
    3 \alpha_4 \chi^2 + 8 \alpha_3 (1 + \chi))}{2 \ell^2 r^2} \\
    0 &=&  \frac{4 \chi (12 L^2 + 6 \alpha_2 + 4 \alpha_3 + 
   4 \alpha_3 \chi + \alpha_4 \chi^2)}{\ell^4}
\end{eqnarray}
In the end, we found the exact AdS solution to the Horndeski theory with linear dilaton field with, 
\begin{equation}
    \alpha_1 =  \frac{1}{6} (-\alpha_3 \chi^2 (1 + \chi) + 
    3\ell^2 (1 - 5 \ell^2)) ~~,~~ \alpha_2 = -8 \ell^2 - 
  \frac{2}{3} \alpha_3 (2 + \chi) ~~,~~ \alpha_4 = (
 4 (9 \ell^2 + \alpha_3))/\chi^2
\end{equation} Black hole solutions are studied in similar classes of theories previously. And numerical black solution with linear dilaton profile will appear soon elsewhere. We have shown that the Lovelock-Horndeski theory indeed contains linear dilaton, thus validating the Fefferman-Graham expansion we implemented in the main letter.

\section{Holographic Conformal Anomaly}

\subsection{Coefficients in Lovelock-Horndeski theory}
We first look at the one without linear dilaton, its anomalous action is given by
\begin{eqnarray}
     \ell^{-1} \mathcal{A} &=&  \alpha_1 \Big( R_{(0)}^2  - 4   R_{(0) ab}   R_{(0)}^{ab} +  R_{(0) abcd}   R_{(0)}^{abcd}  \Big)   +   \frac{ 2 \alpha_1}{\ell^2}  \Big(   R_{(0)}^{ab}   - \frac{1}{2} R_{(0)} g_{(0)}^{ab}  \Big)  g_{(1) ab}          \\
     &+&   A_1  g_{(2)a}^{~~a} +   A_2  g_{(1)ab} g_{(1)}^{~~ab}  +   A_3 \Big(  g_{(1)a}^{~~a} \Big)^2  +   A_4 \Big( \phi_{(1)} \Big)^2     \nonumber   \\
     &+&    A_5  \nabla_a \phi_{(0)}  \nabla^a \phi_{(1)} +   A_6  g_{(1) ab} \nabla^a \phi_{(0)}  \nabla^b \phi_{(0)}   \nonumber  \\
     &+&      \alpha_2 \Big(   R_{(0)}^{ab}   - \frac{1}{2} R_{(0)} g_{(0)}^{ab} \Big)   \nabla_a  \phi^{(0)}  \nabla_b \phi_{(0)}   +  \alpha_4 \Big( \nabla \phi_{(0)} \Big)^4       \nonumber \\
&+&  \big( 12 + \frac{\alpha_2}{\ell^2} \big) g_{(1) a}^{~~a}  \Big(  \nabla \phi_{(0)} \Big)^2  +  \alpha_3 \Big(  \nabla \phi_{(0)}  \Big)^2 \Box \phi_{(0)}     \nonumber 
\end{eqnarray}  where the coefficients $A_i$ are given in the appendix. When solving for the AdS solutions, we have obtained $c1 = 24$, as well as $\alpha_2 = -\big( 4 \ell^2 + \frac{2}{3} \alpha_3 \big) $, in this case though, $\alpha_3$ and $\alpha_4$ are left free. Plugging back the results for $\phi^{(1)}$, $g^{(1)}$ and $g^{(0)}_{ab}$, 
Here we list the coefficients in the conformal anomalies we computed. For the non-critical case,
\begin{eqnarray}
  A_1  =  -  \frac{6}{\ell^4} \Big( \ell^2 - 2 \alpha_1 \Big)  &~~,~~&  A_2 =  \frac{2}{\ell^4}  \Big( \ell^2  -  \alpha_1 \Big)    ~\quad,~\quad 
  A_3  =  -  \frac{1}{2 \ell^4} \Big(  \ell^2 + 2\alpha_1 \Big)  \nonumber   \\
  A_4  =  \frac{4}{\ell^4} \Big( c1\ell^2 + 6\alpha_2 \Big)    &~~,~~& 
  A_5  =  2\Big( c1 + \frac{6 \alpha_2}{\ell^2}  \Big)  ~\quad,~\quad  A_6  =  -   \Big( c1 + \frac{4\alpha_2}{\ell^2} \Big)      
\end{eqnarray} 
Varying with respect to $g_{(2)}$ yields $A_1 = 0$, implying $\alpha_1 = \ell^2 / 2$. Variation with respect to $\phi_{(1)}$ gives:
\begin{eqnarray*}
    \partial_a \frac{\delta \mathcal{A}}{\delta \partial^a  \phi_{(1)}}  -  \frac{\delta \mathcal{A}}{\delta \phi_{(1)}} = 
        -32\frac{\lambda}{\ell^4} \phi_{(1)} -  \nabla^a \Big( \frac{8 \lambda}{\ell^2} \nabla_a \phi_{(0)} \Big) = 0 
\end{eqnarray*} and thus
\begin{equation}
    \phi_{(1)} =  \frac{\ell^2}{4} \Box \phi_{(0)}
\end{equation}
and with respect to $g_{(1)ab}$:
\begin{eqnarray}
\label{g1}
    0 &=&  R_{(0)}^{ab} - \frac{1}{2} R_{(0)} g_{(0)}^{ab}  + \frac{2}{\ell^2} g_{(1)cd} g_{(0)}^{ac} g_{(0)}^{bd} - \frac{2}{\ell^2} g_{(1) cd} g_{(0)}^{cd} g_{(0)}^{ab}  + \Big(  8 -\frac{2\lambda}{3 \ell^2} \Big) g_{(0)}^{ab} \Big(  \nabla \phi_{(0)} \Big)^2  -  4 \Big(  2 -\frac{2\lambda}{3 \ell^2} \Big) \nabla^a \phi_{(0)} \nabla^b \phi_{(0)}     \nonumber
\end{eqnarray}
Multiplying both sides with $g_{(0) ab}$, we find that
\begin{eqnarray}
    g_{(1)a}^{~~a} = -\frac{\ell^2}{6} R_{(0)} + 4 \ell^2 \Big(  \nabla \phi_{(0)} \Big)^2
\end{eqnarray} Plug the above expression back into equation \ref{g1}, we obtain
\begin{eqnarray}
    g_{(1)ab}  &= -\frac{\ell^2}{2} R_{(0)ab} + \frac{\ell^2}{12} R_{(0)}g_{(0)ab}  + \frac{\lambda}{3} \big(  \nabla \phi_{(0)} \big)^2 g_{(0)ab}  + 4 \Big( 1 - \frac{\lambda}{3\ell^2} \Big) \nabla_a \phi_{(0)} \nabla_b \phi_{(0)}
\end{eqnarray}
The final form of the anomaly is obtained as equation (16) in the main letter.
\\

For the critical case, we found the anomalous action,
\begin{eqnarray}
     \ell^{-1} \mathcal{A} &=&  \alpha_1 \Big( R_{(0)}^2  - 4   R_{(0)ab}   R_{(0)}^{ab} +  R_{(0)abcd}   R_{(0)}^{abcd}  \Big)   +   4 B_1 \phi_s \phi_{(2)} + B_1 \big(  \phi_{(1)} \big)^2  - \frac{4\beta \phi_s}{\ell^2} \phi_{(1)}  R_{(0)}   \nonumber  \\
     &+& \frac{96\phi_s}{\ell^2} \phi_{(1)}  g_{(1)a}^{~~a}   +   B_2 ~  g_{(1)ab}  \Big(   R_{(0)}^{ab}   - \frac{1}{2} R_{(0)} g_{(0)}^{ab}  \Big)  +   B_3  g_{(1)}^{ab} g_{(1)ab}  +   B_4 \Big(  g_{(1)a}^{~~a} \Big)^2    
      +   B_5  g_{(2)a}^{~~a}      \nonumber   \\
     &+&   \frac{12}{\ell^2} B_1  \nabla_a \phi_{(0)}  \nabla^a \phi_{(1)} 
      +   B_6  g_{(1)ab} \nabla^a \phi_{(0)}  \nabla^b \phi_{(0)} +  \big( 12 +  \frac{\alpha_2}{\ell^2} \big) g_{(1) a}^{~~a}  \Big(  \nabla \phi_{(0)} \Big)^2  +   \alpha_3 \Big(  \nabla \phi_{(0)}  \Big)^2 \Box \phi_{(0)}     \nonumber  \\
     &+&  \alpha_2 \Big(   R_{(0)}^{ab}   - \frac{1}{2} R_{(0)} g_{(0)}^{~~ab} \Big)   \nabla_a \phi_{(0)}  \nabla_b \phi_{(0)}    
      +   \alpha_4 \Big( \nabla \phi_{(0)} \Big)^4  
\end{eqnarray} where the coefficients $B_i$ and the variations are
\begin{eqnarray}
  B_1  &=&   \frac{24}{\ell^4} \Big( 4 \ell^2 + \beta \Big)  ~~,~~  B_2 =   \frac{2}{\ell^2}  \Big( 2\alpha  +  \beta \phi_s^2 \Big)  -1   ~~,~~
  B_3  =  \ell^2\big(  2 - 24 \phi_s^2  \big)  -2 \alpha + 4\beta \phi_s^2   \nonumber  \\
  B_4 &=&  \frac{-\ell^2}{2} \big( 1 - 24 \phi_s^2 \big)  -  \alpha  -  \beta \phi_s^2     ~~,~~
  B_5  =  6 \Big(  \ell^2  \Big(  8 \phi_s^2  - 1  \Big)   + 2\Big(  \alpha - \beta \phi_s^2 \Big)  \Big)  ~~,~~
  B_6  = -  4 \Big( 6 + \frac{\beta }{\ell^2} \Big)      
\end{eqnarray} 
Variation with respect to $g^{(2)}$ yields $\beta = -4\ell^2$, while variation with respect to $\phi^{(1)}$ gives:
\begin{equation}
    g^{(1)a}_{~~~~a} = \frac{-\ell^2}{6} R^{(0)}
\end{equation}
Then vary the action with respect to $g^{(1)}_{ab}$, we find the equation
\begin{eqnarray}
    \frac{\delta \mathcal{A}}{\delta g^{(1)}_{ab}} = 0 &=& \Big( 1 - 40 \phi_s^2 \Big) R^{(0)ab}  -\frac{1}{2} \Big( 1 - 40 \phi_s^2 \Big) R^{(0)} g^{(0)ab}   + \frac{2}{\ell^2} \Big(  1 - 24 \phi_s^2  \Big) g^{(1)ab}  \nonumber  \\
    &-& \frac{2}{\ell^2} \Big( 1 - 24 \phi_s^2    \Big) g^{(1)a}_{~~~a} g^{(0)ab}  +  8 \Big(  \nabla \phi^{(0)} \Big)^2 g^{(0)ab} - 8 \nabla^a  \phi^{(0)} \nabla^b \phi^{(0)}  + \frac{96}{\ell^2} \phi_s \phi^{(1)} g^{(0)}
\end{eqnarray}   Contract this equation with $g^{(0)}_{ab}$, we get
\begin{equation}
    \phi^{(1)} = -\frac{ \ell^2}{24}\phi_s  R^{(0)} - \frac{\ell^2}{16} \frac{1}{\phi_s} \Big( \nabla \phi^{(0)} \Big)^2
\end{equation}
Substituting the expressions for $g^{(1)}_{ab}$ and $\phi^{(1)}$ into equation (XX), we find 
\begin{eqnarray}
    \frac{2}{\ell^2} \big( 1 - 24 \chi \big) g^{(1)ab} &=& \Big(  \frac{1}{6} - 8\chi^2 \Big) R^{(0)} g^{(0)ab} - \Big( 1 - 40 \chi^2 \Big) R^{(0)ab}  +  8 \nabla^a \phi^{(0)} \nabla^b \phi^{(0)}  -   2 \Big( \nabla \phi^{(0)} \Big)^2 g^{(0)ab}
\end{eqnarray}
Finally, plugging these back into the action, we obtain the holographic conformal anomaly (18) in the main letter.

\subsection{Coefficients for EdGB theory} 
\label{Appendix B}
For theory (5) in four dimensions, the anomalous action takes the form:
\begin{eqnarray}
    \ell^{-1}\mathcal{A} &=&    \Big( \alpha R_{(0)}^{2}  - 4 \alpha R_{(0)ab}   R_{(0)}^{ab} +  \alpha R_{(0)abcd}   R_{(0)}^{abcd}  \Big) e^{-\gamma \phi_{(0)}}  - \Big( \frac{108 \alpha \gamma}{\ell^4} e^{-\gamma \phi_{(0)}} + \lambda  \Lambda e^{\lambda \phi_{(0)}} \Big) \phi_{(2)} \nonumber \\
    &+& \frac{4 \alpha}{\ell^2} \Big(   R_{(0)}^{ab} g_{(1)ab}  - \frac{1}{2} R_{(0)} g_{(0)}^{ab} g_{(1)ab}  \Big) e^{-\gamma \phi_{(0)}}   + \frac{12 \alpha \gamma  \phi_{(1)}   R_{(0)}}{\ell^2} e^{-\gamma \phi_{(0)}}     \nonumber  \\
    &+& \Big( \frac{4}{\ell^2} +  \frac{54 \alpha \gamma^2}{\ell^4} e^{-\gamma \phi_{(0)}} - \frac{\lambda^2 \Lambda}{2} e^{\lambda \phi_{(0)}} \Big) \phi_{(1)}^2  +  2 \nabla_a \phi_{(0)} \nabla^a \phi_{(1)} - g_{(1)ab} \nabla^a \phi_{(0)} \nabla^b \phi_{(0)}     \nonumber \\
    &+&  \Big( - \frac{6}{\ell^2} - \frac{18 \alpha}{\ell^4} e^{-\gamma \phi_{(0)}} - \frac{\Lambda}{2} e^{\lambda \phi_{(0)}} \Big) g_{(2)a}^{~~a}   + \Big( \frac{18 \alpha \gamma}{\ell^4} e^{-\gamma \phi_{(0)}} - \frac{\lambda \Lambda}{2} e^{\lambda \phi_{(0)}}  \Big)  \phi_{(1)}   g_{(0)}^{ab} g_{(1)ab}    \nonumber  \\
    &+&  \Big( \frac{2}{\ell^2} + \frac{7 \alpha }{\ell^4} e^{-\gamma \phi_{(0)}} + \frac{\Lambda}{4} e^{\lambda \phi_{(0)}}  \Big)  g_{(0)}^{ca} g_{(0)}^{db} g_{(1)cd}  g_{(1)ab}     \nonumber  \\
    &+&   \Big( \frac{-1}{2\ell^2} -   \frac{5 \alpha}{2\ell^4} e^{-\gamma \phi^{(0)}} - \frac{\Lambda}{8}  e^{\lambda \phi^{(0)}} \Big) \Big( g_{(0)}^{ab} g_{(1)ab} \Big)^2  
\end{eqnarray}
Variation of the On-shell anomalous action $\mathcal{A}$ with respect to $\phi_{(2)}$.
\begin{eqnarray*}
     0 = \frac{108 \alpha \gamma}{\ell^4} e^{-\gamma \phi_{(0)}} + \lambda  \Lambda e^{\lambda \phi_{(0)} }  \Rightarrow  \Lambda  =  -\frac{108 \alpha }{\ell^4} \frac{\gamma}{\lambda} e^{-(\gamma + \lambda) \phi_{(0)}}
\end{eqnarray*} Then the variation $\frac{\delta \mathcal{A}}{\delta g_{(2)a}^{~~a}}$ give us
\begin{eqnarray*}
     0 = \frac{6}{\ell^2} + \frac{18 \alpha}{\ell^4} e^{-\gamma \phi_{(0)}} + \frac{\Lambda}{2} e^{\lambda \phi_{(0)}}   \Rightarrow
     \frac{1}{\ell^2} = \Big( \frac{9\alpha \eta}{\ell^4} - \frac{3\alpha}{\ell^4} \Big)  e^{-\gamma \phi_{(0)}}   =   \Big( 3 \eta - 1 \Big) \frac{3\alpha}{\ell^4} e^{-\gamma \phi_{(0)}}
\end{eqnarray*}
Defining $\eta \equiv \gamma/\lambda$, we find: $
    e^{-\gamma \phi_{(0)}} = \frac{\ell^2}{3\alpha ( 3\eta - 1 )} $.
With $\eta = 1$, the effective AdS scale becomes $-3 / \alpha$, yielding $\Lambda = -\frac{6}{\ell^2}$.

Further variations with respect to $g^{(1)}_{ab}$ and $\phi_{(1)}$ yield:
\begin{eqnarray*}
    \frac{\delta \mathcal{A}}{\delta \phi_{(1)}} &=& \Big( \frac{18 \alpha \gamma}{\ell^4} e^{-\gamma \phi_{(0)}} - \frac{\lambda \Lambda}{2} e^{\lambda \phi_{(0)}}  \Big) g_{(1)a}^{~~a}  +   \Big( \frac{4}{\ell^2} +  \frac{54 \alpha \gamma^2}{\ell^4} e^{-\gamma \phi_{(0)}} - \frac{\lambda^2 \Lambda}{2} e^{\lambda \phi_{(0)}} \Big) \phi_{(1)}  +   \frac{12\alpha \gamma}{ \ell^2} R_{(0)}  e^{-\gamma \phi_{(0)}}     \nonumber  \\
     \frac{\delta \mathcal{A}}{\delta g_{(1)ab}} &=& 2  \Big( \frac{-1}{2\ell^2}  -  \frac{5 \alpha}{2\ell^4} e^{-\gamma \phi_{(0)}} +  \frac{\Lambda}{8}  e^{\lambda \phi_{(0)}} \Big)   g_{(0)}^{ab}  g_{(0)}^{cd}  g_{(1)cd} + \Big( \frac{2}{\ell^2} + \frac{7 \alpha }{\ell^4} e^{-\gamma \phi_{(0)}} + \frac{\Lambda}{4} e^{\lambda \phi_{(0)}}  \Big)  g_{(0)}^{ca} g_{(0)}^{db} g_{(1)cd}   \nonumber  \\
     &+&  \frac{4\alpha}{\ell^2} \Big(  R_{(0)}^{ab}  - \frac{1}{2} R_{(0)} g_{(0)}^{ab}  \Big)  e^{-\gamma \phi_{(0)}}  +  \Big( \frac{18 \alpha \gamma}{\ell^4} e^{-\gamma \phi_{(0)}} - \frac{\lambda \Lambda}{2} e^{\lambda \phi_{(0)}}  \Big)  \phi_{(1)}  g_{(0)}^{ab} 
\end{eqnarray*} 
Solving these yields $g^{(1)}$ and $g^{(2)}$ in terms of $g^{(0)}$ and $R^{(0)}$. Substituting these back into the anomaly expression leads to equation (14) in the main letter:
\begin{eqnarray}
    \mathcal{A} &=& C_1 \mathrm{R}_{(0)ab}  \, \mathrm{R}^{(0)ab}  + C_2 \, \mathrm{R}_{(0)abcd} \mathrm{R}_{(0)}^{abcd}+ C_3 \mathrm{R_{(0)}^2} + C_4 \Big( \nabla \phi_{(0)}  \Big)^2 + C_5 \, \Box \mathrm{R}_{(0)}  \nonumber \\
    &+& C_6 \,  \mathrm{R}_{(0)} \, \Big( \Box \phi \Big) + C_7 \, \Big( \nabla^a \phi \Big) \Big(\nabla_a \mathrm{R}_{(0)} \Big) + C_8 \, \mathrm{R}_{(0)} \, \Big(\nabla \phi_{(0)} \Big)^2+ C_9 (\nabla^a \nabla_a \phi)(\nabla^b \nabla_b \phi)  \nonumber  \\
    &+& C_{10} \Big(\nabla_a \phi \Big) \Big(\nabla^a \phi \Big) \Big(\Box \phi \Big) + C_{11} \, \Big( \nabla_a \phi \Big) \Big(\nabla^b \nabla_b \nabla^a \phi \Big) + C_{12} \Big( \nabla^b \nabla_b \nabla^a \nabla_a \phi \Big)   \nonumber \\
    &+& C_{13} \Big( \nabla^a \phi_{(0)} \Big)\Big( \nabla^b \phi_{(0)} \Big) + C_{14} \Big( \nabla_a \phi \Big) \Big( \nabla^a \phi \Big) \Big( \nabla_b \phi \Big) \Big(\nabla^b \phi \Big)  \nonumber  \\
    &+& C_{15} \Big(\nabla^a \phi \Big) \Big(\nabla_b \nabla_a \phi \Big) \Big(\nabla^b \phi \Big)+ C_{16} \, \mathrm{R}_{(0)}^{ab} \Big(\nabla_b \nabla_a \phi \Big)+ C_{17} (\nabla_b \nabla_a \phi)(\nabla^b \nabla^a \phi)
\end{eqnarray}with coefficients $\{ C_i \}$ are given in the following. The full expressions for coefficients in holographic conformal anomaly of EdGB theory
\begin{equation}
\begin{aligned}
E_1 &: \quad  \frac{ \ell^2 (81 \gamma^2 - 270 \gamma \lambda + 65 \lambda^2) }{12 (9 \gamma - \lambda) \lambda} \mathrm{R}_{(0)ab}  \, \mathrm{R}^{(0)ab} 
\quad \quad 
E_2: \quad   \frac{1}{3} \ell^3\, \mathrm{R}_{(0)abcd} \mathrm{R}_{(0)}^{abcd}  \\
E_3 &: \quad - \frac{\ell^2}{432 (9 \gamma - \lambda) \lambda (-81 \gamma^3 \lambda + \lambda^2 + 3 \gamma \lambda (-4 + 3 \lambda^2) + 27 \gamma^2 (1 + 8 \lambda^2))^2} \\
& \times \Big(  \big(6377292 \gamma^8 \lambda^2 + 732 \lambda^6 - 118098 \gamma^7 \lambda (36 + 444 \lambda^2) + 6 \gamma \lambda^5 (28 (-31 + 21 \lambda^2) + 8 (-316 + 201 \lambda^2))   \\
& \quad +  9 \gamma^2 \lambda^4 (4 (1423 + 2268 \lambda^2 + 441 \lambda^4) + 8 (2251 - 72 \lambda^2 + 603 \lambda^4))    \\
& \quad +  162 \gamma^3 \lambda^3 (8 (-632 - 1601 \lambda^2 + 912 \lambda^4) + 4 (-362 - 2621 \lambda^2 + 1785 \lambda^4))   \\
& \quad +  6561 \gamma^6 (4 (9 + 252 \lambda^2 + 1471 \lambda^4) + 8 (9 + 360 \lambda^2 + 2065 \lambda^4))  \\
& \quad -   4374 \gamma^5 \lambda (4 (33 + 627 \lambda^2 + 2614 \lambda^4) + 8 (60 + 1005 \lambda^2 + 2872 \lambda^4))    \\
& \quad +  243 \gamma^4 \lambda^2 (8280 + 64768 \lambda^2 - 17592 \lambda^4 + 4 (549 + 7448 \lambda^2 + 20373 \lambda^4 ))\big)  \Big) \mathrm{R_{(0)}^2} \\
E_4 &:  \quad + \frac{9 \ell^3\gamma^2 \lambda (-3\gamma + \lambda)^2 \big(\lambda + 9\gamma^2 \lambda + \gamma (-3 + 9 \lambda^2) \big)}{2 \left( -81 \gamma^3 \lambda + \lambda^2 + 3 \gamma \lambda (-4 + 3 \lambda^2) + 27 \gamma^2 (1 + 8 \lambda^2) \right)^2 } \Big( \nabla \phi_{(0)}  \Big)^2 \quad \\
E_5 &: \quad  - \frac{216 \ell^3\gamma^2 (3 \gamma - 5 \lambda) \lambda^2 }{ (9 \gamma - \lambda) (81 \gamma^3 \lambda - \lambda^2 - 27 \gamma^2 (1 + 8 \lambda^2) + \gamma (12 \lambda - 9 \lambda^3)) } \, \Box \mathrm{R}_{(0)}\\
E_6 &: \quad   \frac{ \ell^3\gamma }{6 (9\gamma-\lambda) \left( -81 \gamma^3 \lambda + \lambda^2 + 3\gamma \lambda (-4+3\lambda^2)+27\gamma^2(1+8\lambda^2) \right)^2 (\lambda+9\gamma^2\lambda+\gamma(-3+9\lambda^2))} \\
& \quad \quad \times \Big(2187 \gamma^5 \lambda - 7 \lambda^4 - 243 \gamma^4 (3 + 34 \lambda^2) + 162 \gamma^3 \lambda (7 + 58 \lambda^2)  \\
& \quad \quad  - 18 \gamma^2 \lambda^2 (32 + 117 \lambda^2) + \gamma (114 \lambda^3 - 63 \lambda^5) \Big)  \,  \mathrm{R}_{(0)} \, \Big( \Box \phi \Big) \\
E_7 &: \quad + \frac{3 \ell^3\gamma }{2 (9\gamma-\lambda) (\lambda+9\gamma^2\lambda+\gamma(-3+9\lambda^2)) (81\gamma^3\lambda-\lambda^2-27\gamma^2(1+8\lambda^2)+\gamma(12\lambda-9\lambda^3))} \\
& \quad \quad \times \Big( 2187 \gamma^5 \lambda - 11 \lambda^4 - 243 \gamma^4 (3 + 38 \lambda^2) - 18 \gamma^2 \lambda^2 (46 + 175 \lambda^2) \\
& \quad \quad + 54 \gamma^3 \lambda (27 + 244 \lambda^2) + \gamma (174 \lambda^3 - 99 \lambda^5)  \Big)  \, \Big( \nabla^a \phi \Big) \Big(\nabla_a \mathrm{R}_{(0)} \Big)  \\
E_8  &:  \quad + \frac{ \ell^3 }{72 (9\gamma-\lambda) \lambda \left( -81\gamma^3\lambda+\lambda^2+3\gamma\lambda(-4+3\lambda^2)+27\gamma^2(1+8\lambda^2) \right)^2 }  \\
& \quad \quad \times \Big( 2187 \gamma^5 \lambda - 7 \lambda^4 - 243 \gamma^4 (3 + 34 \lambda^2) + 162 \gamma^3 \lambda (7 + 58 \lambda^2) \\
& \quad \quad  - 18 \gamma^2 \lambda^2 (32 + 117 \lambda^2) + \gamma (114 \lambda^3 - 63 \lambda^5) \Big) \, \mathrm{R}_{(0)} \, \Big(\nabla \phi_{(0)} \Big)^2   \\
E_9  &=  \frac{ \ell^3(3\gamma-\lambda) (9565938 \gamma^{11} \lambda^4 - 29 \lambda^7 - 1594323 \gamma^{10} \lambda^3 (7+30\lambda^2) + 59049 \gamma^9 \lambda^2 (81+663\lambda^2+450\lambda^4) ) }{8 (9\gamma-\lambda) \lambda \left( -81\gamma^3\lambda+\lambda^2+3\gamma\lambda(-4+3\lambda^2)+27\gamma^2(1+8\lambda^2) \right)^2 (\lambda+9\gamma^2\lambda+\gamma(-3+9\lambda^2))^2 } \\
& \quad \times (\nabla^a \nabla_a \phi)(\nabla^b \nabla_b \phi)     \\
E_{10}  &=  - \frac{ \ell^3\gamma (3\gamma-\lambda)^2 (81\gamma^3\lambda-\lambda^2-9\gamma^2(3+16\lambda^2)+\gamma(12\lambda-9\lambda^3)) \Big(\nabla_a \phi \Big) \Big(\nabla^a \phi \Big) \Big(\Box \phi \Big)}{2 (9\gamma-\lambda) \left( -81\gamma^3\lambda+\lambda^2+3\gamma\lambda(-4+3\lambda^2)+27\gamma^2(1+8\lambda^2) \right)^2 (\lambda+9\gamma^2\lambda+\gamma(-3+9\lambda^2))}  
\end{aligned}
\end{equation}

\begin{equation}
    \begin{aligned}
E_{11}  &=  \quad \frac{ \ell^3\, \Big( \nabla_a \phi \Big) \Big(\nabla^b \nabla_b \nabla^a \phi \Big)}{4 (9\gamma-\lambda) \lambda (81\gamma^3\lambda-\lambda^2-27\gamma^2(1+8\lambda^2)+\gamma(12\lambda-9\lambda^3))} \times \\
   & \quad \Big( 4374 \gamma^5 \lambda - 31 \lambda^4 - 6 \gamma \lambda^3 (-82 + 45\lambda^2) + 324 \gamma^3 \lambda (13 + 54 \lambda^2)  - 243 \gamma^4 (9 + 68 \lambda^2) - 18 \gamma^2 \lambda^2 (131 + 186 \lambda^2) \Big)   \\
    E_{12} &=  \frac{9 \ell^3\gamma (9\gamma^2 - 18\gamma\lambda + 5\lambda^2)}{81\gamma^3\lambda - \lambda^2 - 27\gamma^2(1+8\lambda^2)+\gamma(12\lambda-9\lambda^3)} \Big( \nabla^b \nabla_b \nabla^a \nabla_a \phi \Big) \\
    E_{13} &= \frac{ \ell^3\mathrm{R}_{(0)ab} \Big( \nabla^a \phi_{(0)} \Big)\Big( \nabla^b \phi_{(0)} \Big) }{4 (9\gamma-\lambda) \lambda (81\gamma^3\lambda-\lambda^2-27\gamma^2(1+8\lambda^2)+\gamma(12\lambda-9\lambda^3))}   \times  \\
&  \Big( -4374 \gamma^5 \lambda + 23 \lambda^4 + 6 \gamma \lambda^3 (-62+33\lambda^2) - 324 \gamma^3 \lambda (11+60\lambda^2) + 243 \gamma^4 (9+76\lambda^2) + 18 \gamma^2 \lambda^2 (103+222\lambda^2) \Big) \\
E_{14}   &= - \frac{ \ell^3(3\gamma-\lambda)^2 (81\gamma^3\lambda-\lambda^2-9\gamma^2(3+16\lambda^2)+\gamma(12\lambda-9\lambda^3)) }{48 (9\gamma-\lambda) \lambda \left( -81\gamma^3\lambda+\lambda^2+3\gamma\lambda(-4+3\lambda^2)+27\gamma^2(1+8\lambda^2) \right)^2 }  \Big( \nabla_a \phi \Big) \Big( \nabla^a \phi \Big) \Big( \nabla_b \phi \Big) \Big(\nabla^b \phi \Big) \\
E_{15}   &= - \frac{3 \ell^3\gamma (3\gamma-\lambda)^2 }{81\gamma^3\lambda - \lambda^2 - 27\gamma^2(1+8\lambda^2) + \gamma(12\lambda-9\lambda^3)} \Big(\nabla^a \phi \Big) \Big(\nabla_b \nabla_a \phi \Big) \Big(\nabla^b \phi \Big) \\
E_{16} &=  \frac{9 \ell^3\gamma (9\gamma^2 - 18\gamma\lambda + 5\lambda^2)}{(9\gamma-\lambda)(\lambda+9\gamma^2\lambda+\gamma(-3+9\lambda^2))} \, \mathrm{R}_{(0)}^{ab} \Big(\nabla_b \nabla_a \phi \Big) \\
E_{17} &=  \frac{3 \ell^3(59049 \gamma^9 \lambda^3 - 5 \lambda^6 - 27 \gamma \lambda^5 (-4+5\lambda^2) - 6561 \gamma^8 \lambda^2 (9+8\lambda^2) - 2187 \gamma^7 \lambda (-9-32\lambda^2+69\lambda^4) }{4 (9\gamma-\lambda)\lambda (\lambda+9\gamma^2\lambda+\gamma(-3+9\lambda^2))^2 (81\gamma^3\lambda-\lambda^2-27\gamma^2(1+8\lambda^2)+\gamma(12\lambda-9\lambda^3))} \\
& \quad \times (\nabla_b \nabla_a \phi)(\nabla^b \nabla^a \phi) \\
    \end{aligned}
\end{equation}

\section{Kaluza-Klein Reduction of Gauss-Bonnet and Horndeski Terms}
\label{Appendix A}
In this Appendix, we present the details of KK reductions.\\ 
First, we note that the higher dimensional metric $\hat{g}$ decomposes as 
\begin{equation}
    \hat{g}_{MN} = e^{-2\alpha \phi } g_{ab} \bigoplus e^{-2\beta \phi } g_{ij}
\end{equation} And we also have 
\begin{equation}
    \sqrt{-\hat{g}} = \sqrt{-(e^{-2\alpha \phi })^d (e^{-2\beta \phi })^n g} = e^{-2(d\alpha + n\beta) \phi}\sqrt{-g}
\end{equation}
The dilaton field  descends down directly after the KK reduction,
\begin{equation}
    \hat{\Phi}(x, x') = \Sigma_{n=0}^{\infty} \Phi(x) e^{2n\pi i x'} \approx \Phi_0(x) = \phi(x) 
\end{equation} Thus we assume that the dilaton field depends only on the external space coordinates, $\hat{\Phi} = \phi(x)$. This then gives,
\begin{eqnarray}
    \hat{g}^{MN}\nabla_{M} \hat{\Phi} \nabla_N \hat{\Phi}= e^{-2\alpha \phi} g^{ab} \nabla_a \hat{\Phi} \nabla_b  \hat{\Phi}  
\end{eqnarray}
For the scalar curvature term, $\hat{R}^2$, we have
\begin{eqnarray*}
    \hat{R}^2 &=& e^{-4 \beta  \phi} \lambda^2 +  2 e^{-2 \alpha \phi -  2 \beta  \phi} \lambda R +  e^{-4 \alpha \phi} R^2 -  4 e^{-2 \alpha \phi - 2 \beta  \phi} \big( (-1 + d) \alpha + 
    n \beta \big) \lambda (\nabla_e \nabla^e \phi) \\
    &-&  4 e^{-4 \alpha \phi} \big( (-1 + d) \alpha +  n \beta \big) R (\nabla_e \nabla^e \phi) -  2 e^{-2 \alpha \phi - 2 \beta  \phi} \big( (2 - 3 d +        d^2  \big) \alpha^2  \\
    &+&   2 (-2 + d) n \alpha\beta +  n (1 + n) \beta^2) \lambda (\nabla_e \phi) (\nabla^e \phi)  -  2 e^{-4 \alpha \phi} ((2 - 3 d +   d^2) \alpha^2 +   2 (-2 + d) n \alpha\beta \\
    &+&  n (1 + n) \beta^2) R (\nabla_e \phi) (\nabla^e \phi)  +  4 e^{-4 \alpha \phi} ((-1 + d) \alpha +  n \beta)^2 (\nabla_e \nabla^e \phi) (\nabla_f \nabla^f \phi) \\
    &+&  4 e^{-4 \alpha \phi} \Big( (-2 + d) (-1 + d)^2 \alpha^3 +   3 (2 - 3 d + d^2) n \alpha^2 \beta \\
    && + n (-1 + d - 5 n +  3 d n) \alpha\beta^2  +  n^2 (1 + n) \beta^3 \Big) (\nabla_e \phi) (\nabla^e \phi) (\nabla_f \nabla^f \phi) \\
    &+&  e^{-4 \alpha \phi} \big( (2 - 3 d +  d^2) \alpha^2 +   2 (-2 + d) n \alpha\beta  + n (1 + n) \beta^2  \big)^2 (\nabla_e \phi) (\nabla^e \phi) (\nabla_f \phi) (\nabla^f \phi)
\end{eqnarray*} where we have used symbol $\lambda$ to denote the constant curvature of the internal space.
And for the Ricci tensor squared term, we have
\begin{eqnarray*}
\hat{R}^{MN} \hat{R}_{MN}  =    \hat{R}^{ab} \hat{R}_{ab}  +   \hat{R}^{ij} \hat{R}_{ij}    
\end{eqnarray*} where 
    \begin{eqnarray*} 
        \hat{R}^{ij} \hat{R}_{ij} &=& e^{-4 \beta \phi }  \lambda_{ef}   \lambda^{ef} - 2 e^{-2 \alpha \phi  - 2 \beta \phi} \beta \lambda (\nabla_e \nabla^e \phi ) \\
        &-&  2 e^{-2 \alpha \phi  -  2 \beta \phi }  \beta ((-2 + d) \alpha +  n \beta) \lambda (\nabla_e \phi ) (\nabla^e \phi ) +  e^{-4 \alpha \phi }  n \beta^2 (\nabla_e \nabla^e \phi ) (\nabla_f \nabla^f \phi ) \\
        &+&  2 e^{-4 \alpha \phi }    n \beta^2 ((-2 + d) \alpha + n \beta) (\nabla_e \phi ) (\nabla^e \phi ) (\nabla_f \nabla^f \phi ) \\
        &+&  e^{-4 \alpha \phi }   n \beta^2 ((-2 + d) \alpha + n \beta)^2 (\nabla_e \phi ) (\nabla^e \phi ) (\nabla_f \phi ) (\nabla^f \phi )                   
    \end{eqnarray*}

\begin{eqnarray*}
        \hat{R}^{ab} \hat{R}_{ab} &=&  e^{-4 \alpha \phi}  R_{ef}   R^{ef} -  2 e^{-4 \alpha \phi} \alpha R (\nabla_e \nabla^e \phi ) -   2 e^{-4 \alpha \phi} \alpha ((-2 + d) \alpha +  n \beta ) R (\nabla_e \phi ) (\nabla^e \phi ) \\
        &+&   e^{-4 \alpha \phi} \alpha ((-4 +  3 d) \alpha + 2 n \beta ) (\nabla_e \nabla^e \phi ) (\nabla_f \nabla^f \phi ) \\
        &+&  2 e^{-4 \alpha \phi} \alpha \big(  (6 - 7 d + 2 d^2) \alpha^2 +  3 (-2 + d) n \alpha \beta   + n (1 + n) \beta ^2 \big) (\nabla_e \phi ) (\nabla^e \phi ) (\nabla_f \nabla^f \phi ) \\
        &+&  e^{-4 \alpha \phi} (2 (-2 + d) \alpha^2 +  4 n \alpha \beta  - 2 n \beta ^2)  R_{ef}  (\nabla^e \phi ) (\nabla^f \phi ) \\
        &+&  e^{-4 \alpha \phi} ((-2 + d)^2 (-1 + d) \alpha^4 + 2 (2 - 3 d + d^2) n \alpha^3 \beta  \\
        && + d n^2 \alpha^2 \beta ^2 - 2 n^2 \alpha \beta ^3 + n^2 \beta ^4) (\nabla_e \phi ) (\nabla^e \phi ) (\nabla_f \phi ) (\nabla^f \phi ) \\
        &+&  2 e^{-4 \alpha \phi} \Big( -(-2 + d)^2 \alpha^3 -  3 (-2 + d) n \alpha^2 \beta  \\
        && +   n (d - 2 (1 + n)) \alpha \beta ^2 + n^2 \beta ^3 \Big) (\nabla^e \phi ) (\nabla_f \nabla_e \phi ) (\nabla^f \phi ) \\
        &-&  2 e^{-4 \alpha \phi} ((-2 + d) \alpha +   n \beta )  R_{ef}  \nabla^f \nabla^e \phi  +  e^{-4 \alpha \phi} \big( (-2 + d) \alpha +  n \beta \big)^2 (\nabla_f \nabla_e \phi ) (\nabla^f \nabla^e \phi )          
    \end{eqnarray*} 

  Finally for the Riemann curvature squared, we have 
\begin{eqnarray*}
    \hat{R}^{IJMN} \hat{R}_{IJMN} = \hat{R}^{abcd} \hat{R}_{abcd} + \hat{R}^{ibjd} \hat{R}_{ibjd} + \hat{R}^{ijkl} \hat{R}_{ijkl}
\end{eqnarray*} where
\begin{eqnarray*}
    \hat{R}^{abcd} \hat{R}_{abcd} &=&  e^{-4 \alpha \phi}  R_{efij}  R^{efij} 
         -4 e^{-4 \alpha \phi} \alpha^2 R (\nabla_e \phi) (\nabla^e\phi) \\
        &+& 4 e^{-4 \alpha \phi} \alpha^2 (\nabla_e \nabla^e\phi ) (\nabla_f \nabla^f\phi ) +  8 (-2 + d) e^{-4 \alpha \phi} \alpha^3 (\nabla_e\phi) (\nabla^e\phi) ( \nabla_f \nabla^f\phi ) \\
        &+& 8 e^{-4 \alpha \phi} \alpha^2  R_{ef}  (\nabla^e \phi) (\nabla^f \phi) - 8 e^{-4 \alpha \phi} \alpha R_{ef}  \nabla^f \nabla^e\phi  \\
        &+&  2 (2 - 3 d +   d^2) e^{-4 \alpha \phi} \alpha^4 (\nabla_e \phi) (\nabla^e \phi) (\nabla_f \phi) (\nabla^f \phi) \\
        &-&  8 (-2 + d) e^{-4 \alpha \phi} \alpha^3 (\nabla^e \phi) (\nabla_f \nabla_e \phi ) (\nabla^f \phi) \\
        &+&  4 (-2 + d) e^{-4 \alpha \phi} \alpha^2 (\nabla_f \nabla_e \phi) (\nabla^f \nabla^e\phi )           \\
    \hat{R}^{ibjd} \hat{R}_{ibjd} &=&  8 d e^{-4 \alpha \phi} \alpha \beta^2  
        ( \nabla_e \phi) (\nabla^e\phi) (\nabla_f \nabla^f\phi)   \\  
        &+&  4 d e^{-4 \alpha \phi} \beta^2 (d \alpha^2 + \beta (-2 \alpha + \beta) ) (\nabla_e \phi) (\nabla^e \phi) (\nabla_f \phi) (\nabla^f \phi)    \\
        &+&  8 d e^{-4 \alpha \phi} \beta^2 (-2 \alpha + \beta) (\nabla^e \phi) (\nabla_f \nabla_e\phi ) (\nabla^f \phi)   + 4 d e^{-4 \alpha \phi} \beta^2 (\nabla_f \nabla_e\phi) (\nabla^f\nabla^e \phi)          \\        
    \hat{R}^{ijkl} \hat{R}_{ijkl} &=&  e^{-4 \beta \phi}  \lambda_{efij}   
         \lambda^{efij} -  4 e^{-2 \alpha \phi -  2 \beta \phi} \beta^2 \lambda (\nabla_e \phi) (\nabla^e \phi) \\
        &+& 2 e^{-4 \alpha \phi} (-1 + n) n \beta^4 (\nabla_e \phi) (\nabla^e \phi) (\nabla_f \phi) (\nabla^f \phi)           
\end{eqnarray*}
Combining the above pieces, we found the reduced Gauss-Bonnet term,
\begin{eqnarray*}
    \hat{\mathcal{L}}_{GB} &=&-4 e^{-4 \alpha \phi} \big( (d-3) \alpha +  n \beta \big) \mathcal{R} \nabla_a \nabla^a \phi +   8 e^{-4 \alpha \phi} \big( (d-3) \alpha +  n \beta \big) \mathcal{R}_{ab} \nabla^a \nabla^b \phi    \\
    &&-2 e^{-4 \alpha \phi} \big( (d-3) (d-4) \alpha^2 + 2 (d-4) n \alpha \beta + n (1 + n) \beta^2 \big) \mathcal{R} \nabla_a \phi \nabla^a \phi   +  \\
    &&+ 8 e^{-4 \alpha \phi} \big( -((d-3) \alpha^2) - 2 n \alpha] \beta + n \beta^2 \big) \mathcal{R}_{ab} \nabla^a \phi \nabla^b \phi    \\
    &&+ 4 e^{-4 \alpha \phi} \big(  (d-3)^2 (d-2) \alpha^3 + 3 (6 - 5 d + d^2) n \alpha^2 \beta   \\
    && \quad  \quad \quad \quad + n (3 - 7 n + d ( 3 n - 1) ) \alpha \beta^2 + (n - 1) n^2 \beta^3 \big) \nabla_a \nabla^a \phi \nabla_b \phi \nabla^b \phi       \\
    && + e^{-4 \alpha \phi} \big( (24 - 50 d + 35 d^2 - 10 d^3 +  d^4) \alpha^4 + 4 (-8 + 14 d - 7 d^2 + d^3) n \alpha^3 \beta + \\
    &&~~~~~~~~~~~  2 (-1 + d) n (6 - 10 n + d (-1 + 3 n)) \alpha^2 \beta^2 +   4 (-1 + n) n (2 + (-2 + d) n) \alpha \beta^3   +   \\
    &&~~~~~~~~~~~  n (2 - n - 2 n^2 + n^3) \beta^4)    \nabla_a  \phi \nabla^a \phi \nabla_b \phi \nabla^b \phi     + \\
    &&+8 e^{-4 \alpha \phi} \big(  (6 - 5 d + d^2) \alpha^3 +  3 (-2 + d) n \alpha^2 \beta +   n (d + 2 n) \alpha 
 \beta^2 - (-1 + n) n \beta^3 \big)     \\
 &&~~~~~~~~~~~\nabla^a  \phi \nabla_a \nabla_b \phi \nabla^b \phi       \\
    &&+4 e^{-4 \alpha \phi} \big(  (6 - 5 d + d^2) \alpha^2 + 2 (-2 + d) n \alpha \beta + (-1 + n) n \beta^2 \big)   \nabla_a \nabla^a \phi  \nabla_b \nabla^b \phi   \\
    &&-4 e^{-4 \alpha \phi} \big(  (6 - 5 d + d^2) \alpha^2 +  2 (-2 + d) n \alpha \beta + (-1 + n) n \beta^2 \big)     \nabla_a \nabla_b \phi  \nabla^a \nabla^b \phi 
\end{eqnarray*}
Use the identities $(\Box \phi)^2 - \nabla_a \nabla_b \phi \nabla^a \nabla^b \phi  =  \mathcal{R}^{ab} \nabla_a \phi \nabla_b \phi + \nabla_a ( \Box \phi \nabla^a \phi - \nabla^a \nabla^b \phi \nabla_b \phi ) $ and perform many integration by parts and throwing several total derivative terms, we found that,
\begin{eqnarray}
    \hat{\mathcal{L}}_{GB} &=&  e^{-4\alpha}\Big(  -4 \big( (-4 + d) (-3 + d) \alpha^2 + 
   2 (-3 + d) n \alpha \ + (-1 + n) n \beta^2)  \big) \mathcal{G}^{\mu\nu} \nabla_{\mu} \phi \nabla_{\nu} \phi    \nonumber \\ 
   &&  \quad \quad \quad + \big( -2 (-4 + d) (-3 + d) (-2 + d) \alpha^3 - 
 6 (-3 + d) (-2 + d) n \alpha^2 \beta    \nonumber  \\
 && \quad \quad \quad \quad \quad \quad -  6 (-2 + d) (-1 + n) n \alpha \beta^2 - 2 (-2 + n) (-1 + n) n \beta^3 \big)  (\nabla \phi)^2 \Box \phi    \nonumber   \\ 
 && \quad \quad \quad + \big( -((-4 + d) (-3 + d)^2 (-2+d) \alpha^4) - 4 (-3 + d)^2 (-2 + d) n \alpha^3 \beta   \nonumber  \\
 &&  \quad \quad \quad \quad \quad \quad-  2 (-2 + d) n (3 - 2 (-1 + d) - 5 n + 3 (-1 + d) n) \alpha^2 \beta^2   \nonumber  \\
 && \quad \quad \quad \quad \quad \quad- 4 (-2 + d) (-1 + n)^2 n \alpha \beta^3 - (-2 + n) (-1 + n)^2 n \beta^4 \big) 
 ((\nabla \phi)^2 )^2\Big)     \nonumber  \\
    &+& 2 e^{-2 \alpha \phi - 2 \beta \phi}  \Big(  \lambda \mathcal{R} + \lambda \big( (d-1)(d-2)\alpha^2 + 2(d-1)(n-2) \alpha \beta + (n-2)(n-3) \beta^2 \big) (\nabla \phi)^2 \Big) \nonumber  \\
    &+& e^{-4\alpha} \big( \mathcal{R}^2 - 4 \mathcal{R}^{\mu\nu} \mathcal{R}_{\mu\nu} + \mathcal{R}^{\mu\nu\rho\sigma} \mathcal{R}_{\mu\nu\rho\sigma} \big)  + e^{-4\beta} \big( \mathcal{\lambda}^2 - 4 \mathcal{\lambda}^{\mu\nu} \mathcal{\lambda}_{\mu\nu}    + \mathcal{\lambda}^{\mu\nu\rho\sigma} \mathcal{\lambda}_{\mu\nu\rho\sigma} \big) \nonumber  
\end{eqnarray} For the case we focus where we have set $\alpha = 0$ and $\lambda = 0$ in the reduction ansatz. We found the Horndeski term,
\begin{eqnarray}
    \hat{G}^{\mu\nu} \nabla_{\mu} \Phi \nabla_{\nu} \Phi &=& \mathcal{G}^{\mu\nu} \nabla_{\mu} \phi \nabla_{\nu} \phi + \frac{3}{2} n\beta \Box (\nabla \phi)^2 \phi - \frac{1}{2} n (n-3) \beta^2 ((\nabla \phi)^2)^2
\end{eqnarray} 
Eventually, we found that coefficients for theory (7) in the main letter,
\begin{eqnarray}
    \alpha_1 =  \Tilde{\alpha} ~~&,&~~
    \alpha_2 =  \Tilde{\alpha} \big( a_1 - 4n(n-1) \beta^2 \big)   \quad , \quad 
    \alpha_3 =  \Tilde{\alpha} \big( \frac{3}{2} a_1 n\beta -2n (n-1)(n-2) + a_3 \big) \nonumber  \\ 
    \alpha_4  &=&  \Tilde{\alpha} \big( n(n-1)^2(n-2)\beta^4 - \frac{a_1}{2} n (n-3)\beta^3 + a_4 \big)    
\end{eqnarray}

\end{document}